\documentclass[manuscript]{acmart}
\AtBeginDocument{%
  }
\usepackage{svg}
\usepackage{amsmath}
\usepackage{multirow}
\usepackage{subcaption}

\begin{document}

\title{Multimodal LLM Augmented Reasoning for Interpretable Visual Perception Analysis}
\begingroup
\renewcommand\thefootnote{}\footnote{
 This paper was presented at the 2025 ACM Workshop on Human-AI Interaction for Augmented Reasoning (AIREASONING-2025-01). This is the authors’ version for arXiv.}
\endgroup


\author{Shravan Chaudhari}
\authornotemark[1]
\email{schaud35@jh.edu}
\affiliation{%
  \institution{Amazon.com, Inc.}
  \city{Palo Alto}
  \state{California}
  \country{USA}
  \institution{Johns Hopkins University}
  \city{Baltimore}
  \state{Maryland}
  \country{USA}
}

\author{Trilokya Akula}
\affiliation{%
  \institution{Amazon.com, Inc.}
  \city{Palo Alto}
  \state{California}
  \country{USA}}
\email{atriloky@amazon.com}

\author{Yoon Kim}
\affiliation{%
  \institution{Amazon.com, Inc.}
  \city{Palo Alto}
  \state{California}
  \country{USA}
}
\email{yoonskim@amazon.com}


\author{Tom Blake}
\affiliation{%
  \institution{Amazon.com, Inc.}
  \city{Palo Alto}
  \state{California}
  \country{USA}
}
\email{blaktho@amazon.com}







\begin{abstract}
  In this paper, we advance the study of AI-augmented reasoning in the context of Human-Computer Interaction (HCI), psychology and cognitive science, focusing on the critical task of visual perception. Specifically, we investigate the applicability of Multimodal Large Language Models (MLLMs) in this domain. To this end, we leverage established principles and explanations from psychology and cognitive science related to complexity in human visual perception. We use them as guiding principles for the MLLMs to compare and interprete visual content. Our study aims to benchmark MLLMs across various explainability principles relevant to visual perception. Unlike recent approaches that primarily employ advanced deep learning models to predict complexity metrics from visual content, our work does not seek to develop a mere new predictive model. Instead, we propose a novel annotation-free analytical framework to assess utility of MLLMs as cognitive assistants for HCI tasks, using visual perception as a case study. The primary goal is to pave the way for principled study in quantifying and evaluating the interpretability of MLLMs for applications in improving human reasoning capability and uncovering biases in existing perception datasets annotated by humans.  
\end{abstract}






\maketitle

\section{Introduction}
With the advent of Multimodal Large Language Models (MLLMs), the research in AI-augmented reasoning and explainability has gained momentum. Using LLMs or MLLMs as reasoning agents to aid cognitive tasks such as crowdsourcing or dataset annotation can avoid human biases by providing reat time feedback. In order to entrust MLLMs with such crucial cognitive tasks, it is necessary to validate their reasoning skills. Such validation also makes AI systems (populalrly known to be black box systems) interpretable to downstream application stakeholders like domain experts with limited experience and knowledge in machine learning. In this paper, we align our work with the workshop's research objectives by proposing a novel analytical framework to evaluate and quantify multimodal LLM reasoning to shed light on their applicability in AI-augmented HCI systems. This framework establishes a responsible and explainable LLM benchmarking approach supported by strong empirical validation thereby advancing the evaluation of their utility in reasoning tasks. We test this framework on a key task that lies at the intersection of HCI, cognitive science and psychology, called visual perception analysis, using MLLMs.  \cite{Sun2021CuriousOH,davidson2023complexity&memorability,vangeert2020order,palumbo2014examining}. 

Studying complexity in visual perception has significant real-world applications, including examining how user interface designs, such as advertisements and websites, or product designs, influence users' perception and cognitive load. Furthermore, visually complex images require more time and efforts for a human to perform visual search in or describe it \cite{lin2014microsoft}. Estimation of subjective visual complexity can aid e-commerce services to present the search results in the most optimal way possible by minimizing the cognitive overload of going through all listings for a search query while conveying the important information to potential buyers. 

However, there is no single commonly accepted explanation of visual complexity and there have been several attempts at describing it. Some works base the definition on level of details and intricacies in an image while others refer to the level of difficulty in describing the image as a measure to define visual complexity \cite{heaps1999similarity,snodgrass1980standardized}. \cite{rosenholtz2007measuring} relates visual complexity to image compression and information theory by using visual clutter along with the amount of information conveyed in the image.            \cite{baughan2020websitesvisualcomplexity,Miniukovich2020webpageaesthetics,pieters2010stopping,wu2016complexity,reinecke2013predicting,king2020influence}. Max Wertheimer in \cite{wertheimer1923Gestalt} first introduced Gestalt principles, emphasizing how humans perceive whole forms instead of mere aggregation of their individual parts. Wertheimer's insights into perceptual organization have profound implications for simplifying visual complexity in design. \cite{koffka1935laws} elaborates on the principles by further detailing the impact of proximity, similarity, and closure on our understanding of complex visual stimuli underscoring the importance of these principles in reducing cognitive overload and enhancing clarity in visual communication. \cite{kohler1947Gestalt} provides a holistic discussion of how Gestalt principles guide our interpretation of visual elements and applications of these principles to make information easier to interpret. Recently, gestalt principles have also been applied to simplify complex data visualizations or in UX designs to improve user engagement and comprehension \cite{bradley2019gestalt,cinelli2020gestalt}. 

\section{Contributions}
\begin{itemize}
    \item Our work pioneers reasoning-based assessment by MLLM to compare and evaluate the complexity in visual perception to enhance transparency and trust in AI-driven analyses. Existing deep learning techniques for visual complexity merely learn to map images to human-annotated ratings. In contrast, our approach does not provide an alternative method to predict complexity metric, but instead focuses on quantifying underlying explainable parameters rooted in cognitive and psychology for perception of visual content. Our work seeks to bridge the gap between qualitative and quantitative studies in HCI. 
    \item We present a scalable framework that is not susceptible to annotator biases in a dataset. Deep learning based visual complexity prediction methods that learn or finetune from an annotated training dataset often also learn and exhibit biases prevalent in the dataset which reduces their generalizability to new data. On the other hand, our prompt design constrains the MLLM to assess the images solely on the basis of principles highlighted in the human perception literature.      
    \item We consistently observe a strong high correlation between explainable attributes such as visual clutter and the gestalt principle of simplicity in visual content with complexity in visual perception through rigorous empirical validation on human annotated public datasets (like SAVIOAS, IC9600) across diverse categories. This highlights one of the biases in human cognition while perceiving visual content.  
    
\end{itemize}

\begin{figure}
\centering
  \includegraphics[width=0.75\linewidth]{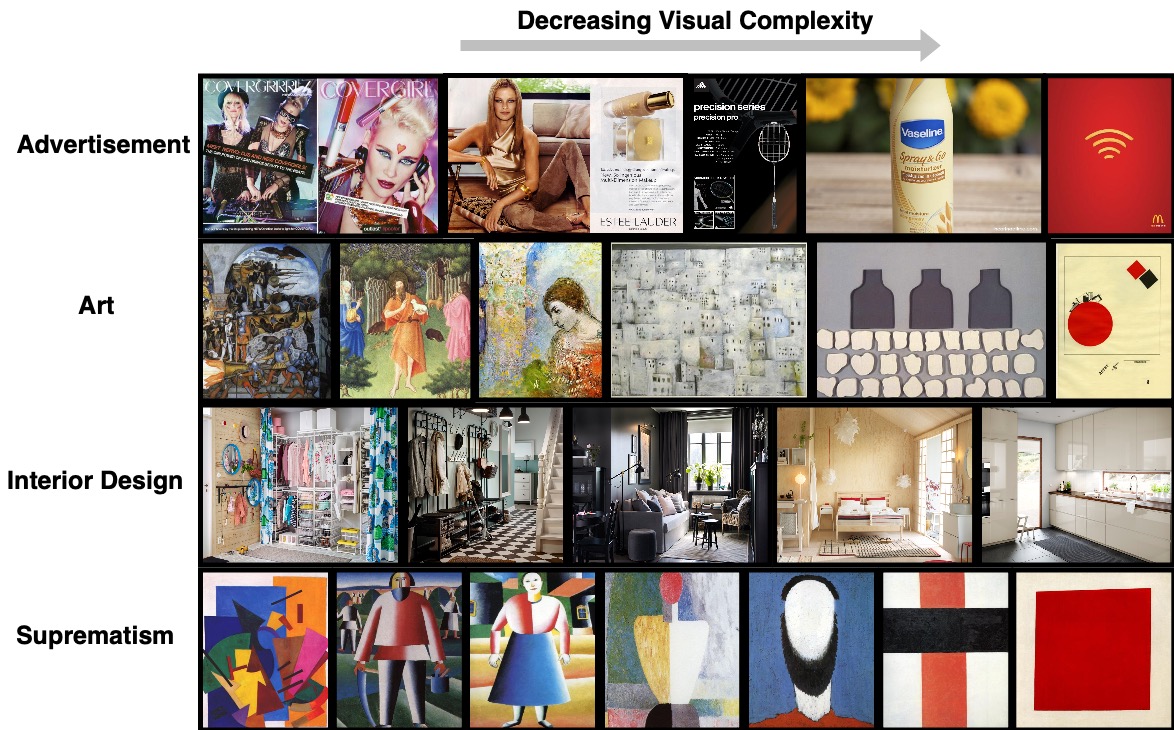}
  \caption{Images ordered based on visual complexity across different categories.}
  \label{fig:intro_fig}
\end{figure}

\section{Related Work}
Some early works such as \cite{chipman1977complexity} demonstrated a positive correlation of visual complexity with the number of elements and a negative correlation with visual symmetry. \cite{fan2017visual} evaluated the visual complexity using features such as color richness, stroke thickness and white spaces. \cite{rosenholtz2007measuring} defined visual complexity using information theory based on the size of the file achieved through JPEG or ZIP compression. Similarly, there have been past studies which use edge density, Fourier slope, entropy and information gain attributes to estimate visual complexity \cite{vangeert2020order}. Deep learning techniques such as \cite{SARAEE2020deepintermediatelayer} used activations from intermediate layers of a CNN pretrained on object or scene recognition tasks which performed best on SAVOIAS dataset which was also introduced by them \cite{saraee2018savoiasdiversemulticategoryvisual}.

Most of the recent advancements in estimation of visual complexity using linear combination of handcrafted features or computer vision techniques fall in the second category since they need some kind of training dataset to learn parameters. \cite{corchs2016predicting} fit a linear combination of 11 handcrafted image features on RSIVL dataset to estimate perceived complexity rating. The features measured spatial, frequency, color, texture, contrast and other image properties. Although such handcrafted features are interpretable, however, such features are dataset specific and are dependent on subjectively defined measures \cite{nath2024simplicity}.  \cite{feng2023ic9600} used a CNN based approach using activation maps to predict the scores which outperformed other baselines on their dataset IC9600 \cite{feng2022ic9600}. However, they used the training dataset of IC9600 to train their model. \cite{nath2024simplicity} proposes a linear combination of handcrafted image features evaluated using foundation models \cite{bommasani2021opportunities} which again requires a training dataset to fit the linear model. \cite{nath2024simplicity} further notes that this method struggles in the presence of images having notable proportion of text content  which is a common occurence in images of advertisement or webpages. \cite{celona24vitforimagecomplexity} leverages a vision transformer to predict image complexity which also requires training dataset to finetune the model. 

LLMs have recently accomplished immense success in understanding, abstracting from and generating text \cite{brown2020languagemodelsarefewshotlearners, Chowdhery2023palm, openai2024gpt4technicalreport, touvron2023llama2openfoundation, zheng2023judging}. This has further lead to a plethora of studies that explore and understand multimodal capabilities of LLMs \cite{Fang2023evaexploringlimitsofmasked,radford2021learningtransferablevisualmodels,sun2023evaclipimprovedtrainingtechniques}. Such models consist of transformer based architectures to parse and attend to multiple input modalities which are pretrained using large image-caption pair dataset \cite{Changpinyo2021Conceptual1P, Krishna2017visualgenome,lin2014microsoft,schuhmann2021laion}. Then these models are finetuned for a particular downstream visual question answering (VQA) task using curated VQA datasets \cite{Antol2015vqa,goyal2019makingvinvqamatter}. Researchers use instruction-following chat data generated by LLMs starting from LLaVA \cite{Liu2023ImprovedBW,liu2024llavanext,li2024llavanext-ablations} in VQA format with the goal of instruction tuning to obtain improved results \cite{bai2023qwenvlversatilevisionlanguagemodel,li2023blip2bootstrappinglanguageimagepretraining,geminiteam2024geminifamilyhighlycapable,Chen2023MiniGPTv2LL}. There have been extensive studies that explore such capabilities of MLLMs further, especially on VQA reasoning \cite{Zellers2018FromRT,schwenk2022aokvqa,fu-etal-2022-theres,fu-etal-2023-generate}. MLLMs have also proven to be efficient in understanding and describing the semantics of input images, pdfs \cite{tang2024pdfchatannotator} etc. This has paved the way for new problems and tasks which can be mostly expressed in natural language utilizing the vision-language connection learned by MLLMs \cite{sable-meyer2022language,alayrac2022flamingovisuallanguagemodel,bai2023qwenvlversatilevisionlanguagemodel,chen2023mllm,chen2023sharegpt4vimprovinglargemultimodal}.

\begin{figure}
  \includegraphics[width=0.65\linewidth]{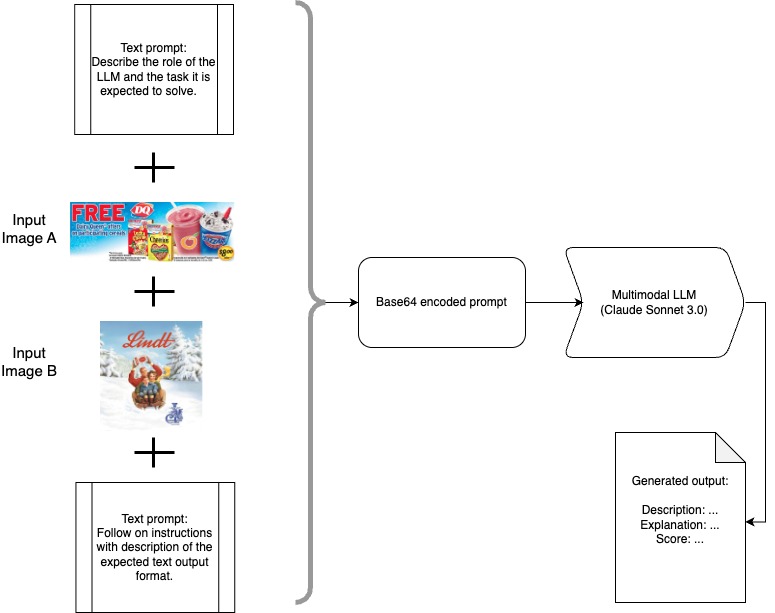}
  \caption{Visual perception analysis framework using multimodal LLMs}
  \label{fig:pipeline}
\end{figure}

\section{Method}
We select 6 fundamental gestalt principles, namely laws of similarity, proximity, simplicity, simplicity, closure, continuity and figure vs. ground, along with the notions of visual clutter and visual symmetry as various explainable parameters of visual perception analysis. We use two public datasets, SAVOIAS \cite{saraee2018savoiasdiversemulticategoryvisual} and IC9600 \cite{feng2022ic9600} consisting of various categories such as advertisement, interior design, paintings, etc. The goal is to ask the MLLM agent to quantify the visual content based on each of these parameters. For the experiments, we primarily use Claude Sonnet 3.0 by Anthropic AI \cite{claudeai} as our MLLM agent. Refer to section \ref{sec:modellimits} for more details on limitations of several MLLM models that were also tried. Next, we need to design appropriate prompt describing the task to the MLLM agent and providing it with the necessary background knowledge to evaluate the provided visual content. 

\subsection{Prompt Design}
\begin{figure}[h!]
  \centering
  \includegraphics[width=0.85\linewidth]{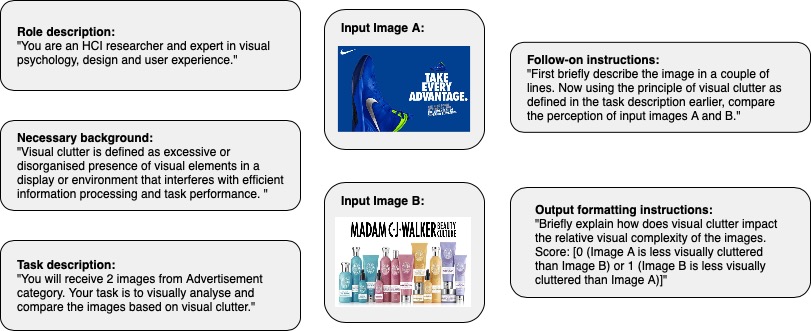} 
  \caption{Various important components of the prompt to compare advertisement images based on visual clutter.}
  \label{fig:prompt_sample}
  \hfill
\end{figure}
We first develop a text prompt to describe the role of the MLLM agent as an HCI researcher whose goal is to evaluate the complexity of the input images. We then provide and define the explainable parameters of evaluating complexity in visual perception as we would to any human annotator. These principles include the 6 gestalt principles along with visual clutter and visual symmetry. A sample prompt is shown in fig. \ref{fig:prompt_sample}. We finally ask the MLLM agent to analyse and compare the input images using each of the 8 explainable parameters of visual perception and provide a sample output format. \\
\cite{zhang2024understanding} suggests that more informative prompts lead to less variability in the LLM responses and further enhance trust in the model's outputs. We use a low temperature parameter of $0.01$ to have less variance in MLLM output. Furthermore, prompt-based rationales are known to align better with human expectations \cite{fayyaz2024evaluating}. Hence, in our prompts, we ask the MLLM to justify their comparative analysis. Asking LLMs for justifications or explanations for their reasoning processes aids in reducing their variability and improve reliability \cite{wang2024reasoning}. 

\subsection{Quantifying reasoning-based visual perception analysis}
Several approaches for rating images have been studied to obtain annotations from humans (or MLLM agents). Evaluating one image at a time faces challenges of biases in the rating scale of annotators \cite{miller1956magical}, while asking annotators to rank each image relative to all others is not feasible beyond 5 to 9 images at a time for humans and beyond the context limit of MLLM agent. It has been demonstrated that pairwise comparison followed by conversion of pairwise ranking to global ranking is a better alternative \cite{arrow1950difficulty,david1963paired,kendall1940method}. Hence, we resort to pairwise comparative analysis based on each explainable parameter. We create a pairwise image comparison matrix for each parameter. Each matrix can be denoted as a count matrix $S=\{s_{i,j}\}$ where $s_{i,j}$ is a binary comparison result based on an explainable parameter for a pair of images $(i,j)$. We finally get the relative score for an explainable parameter of an image, $\hat{s_i} = \frac{1}{n} \sum^{n}_{j=1}s_{i,j}$.   
\subsection{Correlation with complexity in human perception}
Since, we mainly want to capture the correlation between images ranked using explainability parameters and visual complexity scores, we primarily use Pearson correlation (PLCC) \cite{Song2007pearson} and Spearman's rank correlation (SROCC) \cite{GAUTHIER2001spearman} as our metrics to analyze MLLM interpretability. Pearson correlation captures linear relationship while Spearman's correlation is necessary to measure non-linear relationship and monotocity.  

\subsection{MLLM Models}
\label{sec:modellimits}
We evaluated several open-source or publicly available models like LLaVa-NeXT \cite{li2024llavanext-ablations}, DeepSeek-VL \cite{lu2024deepseekvlrealworldvisionlanguageunderstanding} and Llama 3.2 \cite{grattafiori2024llama3herdmodels}. However, we observed that LLaVa-NeXT and DeepSeek-VL were unable to consistently follow the image ordering while explaining their assessment of two images, while Llama 3.2 does not support multiple images in a single prompt making pairwise comparison impossible in the current version.    
For our purpose, we use Claude Sonnet 3.0 which allows larger image sizes of 8000x8000 pixels \cite{claudeai} and allows for multiple input images in the input prompt along with text prompt and text output generation. We use a low temperature parameter of $0.01$ to have less variance in MLLM output. 

\begin{figure}[h!]
  \centering
    \includegraphics[width=0.9\linewidth]{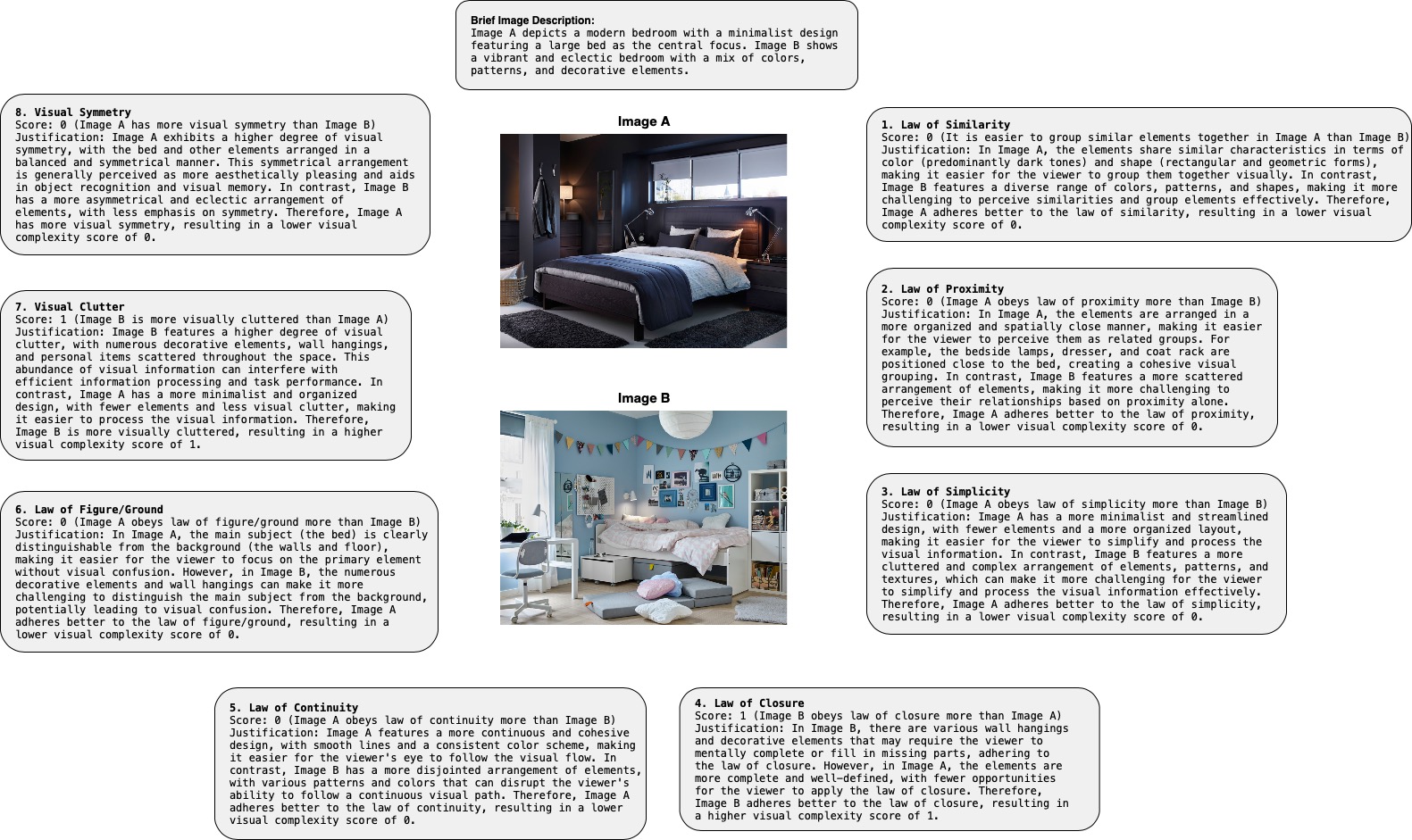} 
    \caption{Sample MLLM (Claude Sonnet 3.0) output comparing two interior design images.}
    \label{fig:MLLM_sample}
\end{figure}

\begin{figure}[t!]
      \centering
      \hspace*{\fill}%
      \begin{subfigure}[t]{0.35\linewidth}  
        \includegraphics[width=\linewidth]{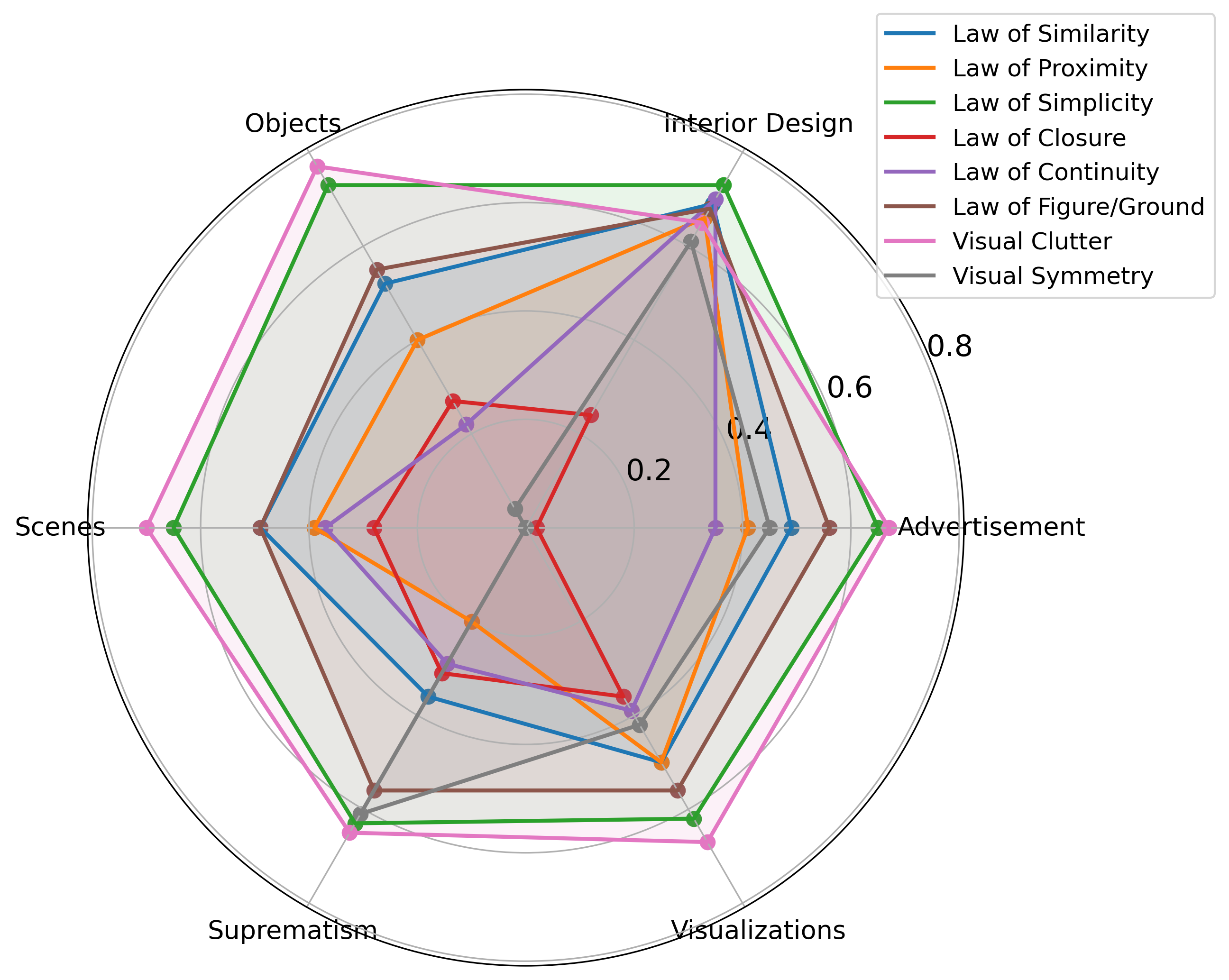} 
        \caption{PLCC scores for SAVIOAS dataset}
        \label{fig:plcc_savioas}
      \end{subfigure}%
      \hfill
      \begin{subfigure}[t]{0.35\linewidth}  
        \includegraphics[width=\linewidth]{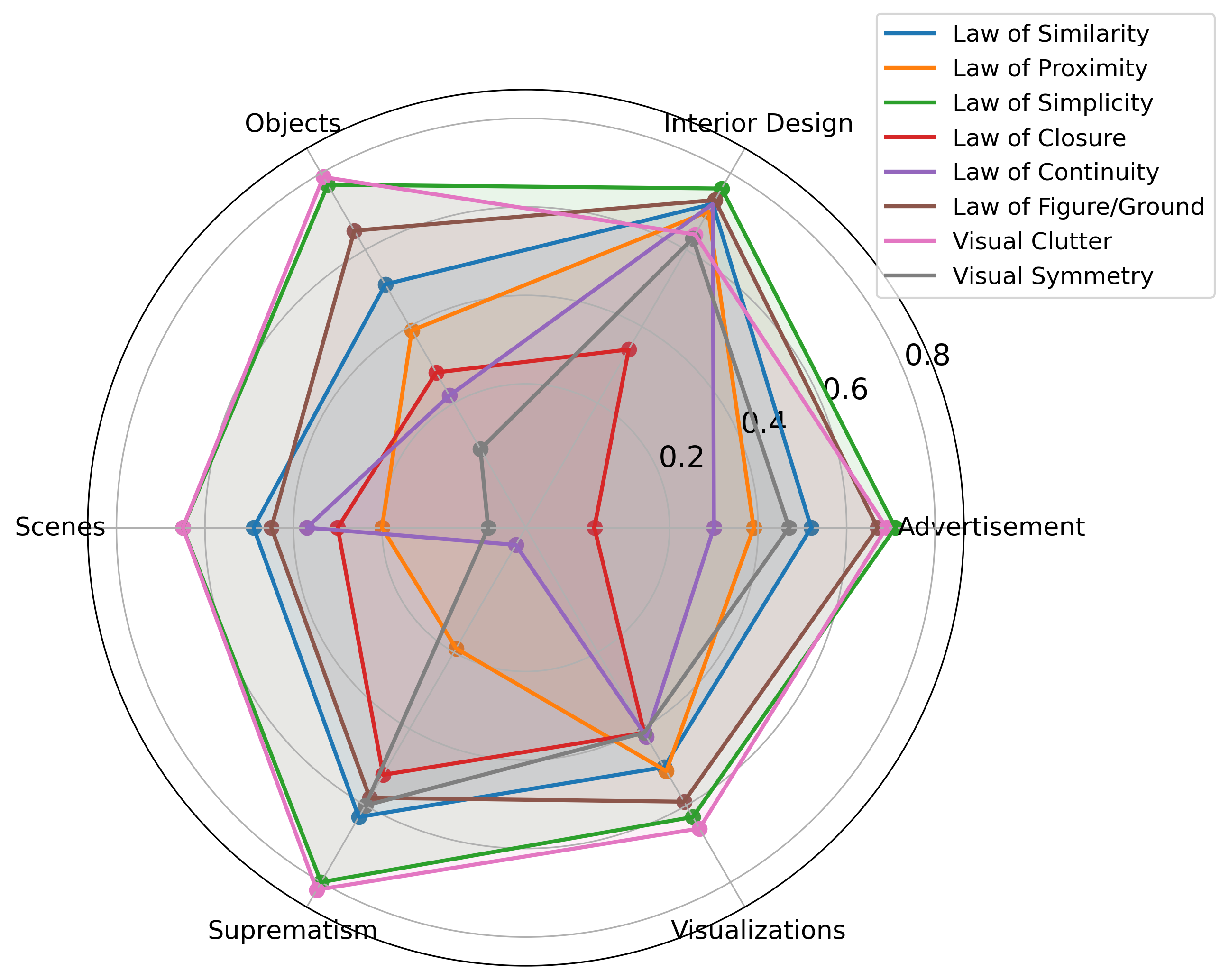} 
        \caption{SROCC scores for SAVIOAS dataset}
        \label{fig:srocc_savioas}
      \end{subfigure}
      \hspace*{\fill}%
      \vfill
      \hspace*{\fill}%
      \begin{subfigure}[t]{0.35\linewidth}  
        \includegraphics[width=\linewidth]{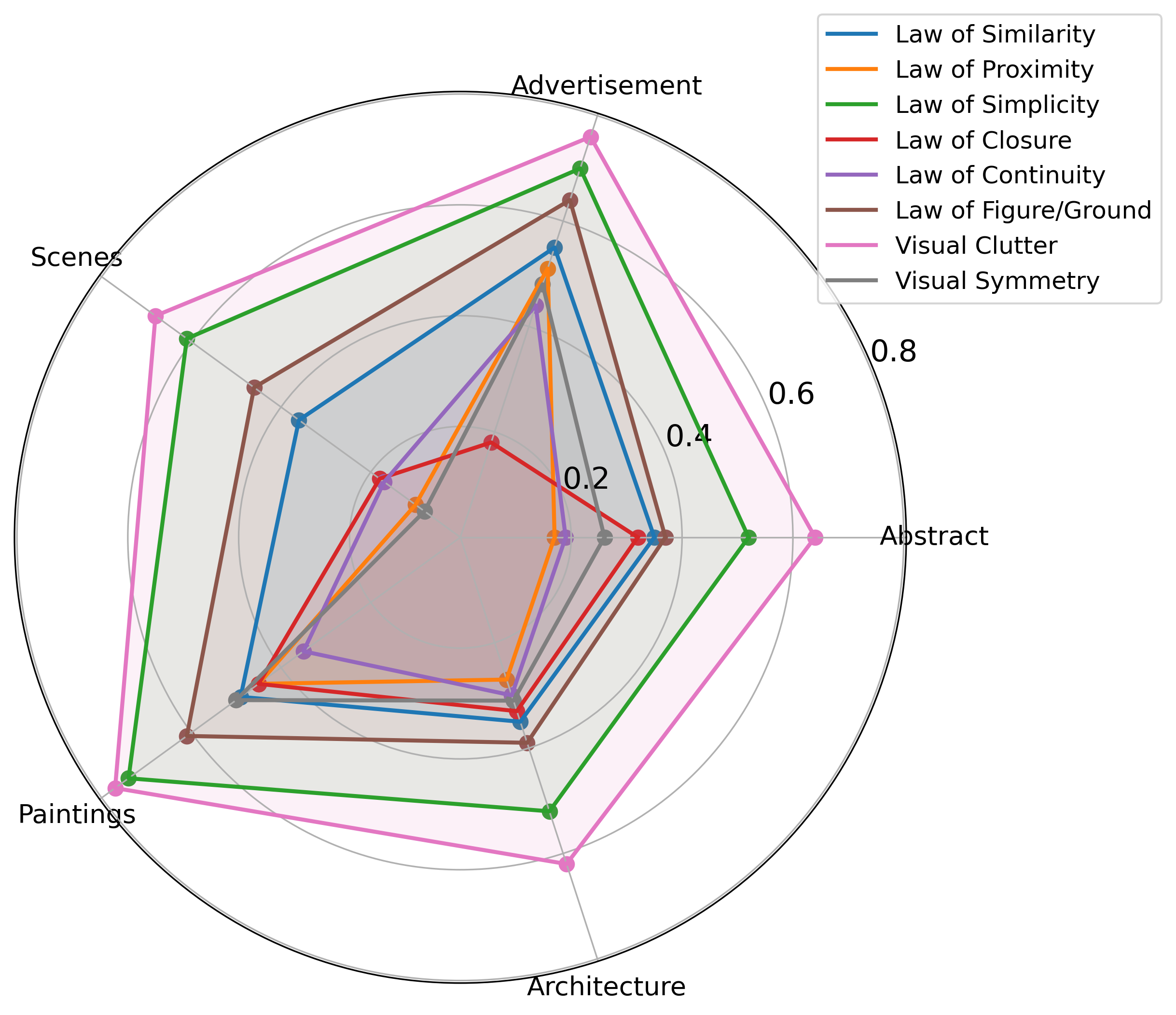} 
        \caption{PLCC scores for IC9600 dataset}
        \label{fig:plcc_ic9600}
      \end{subfigure}%
      \hfill
      \begin{subfigure}[t]{0.35\linewidth}  
        \includegraphics[width=\linewidth]{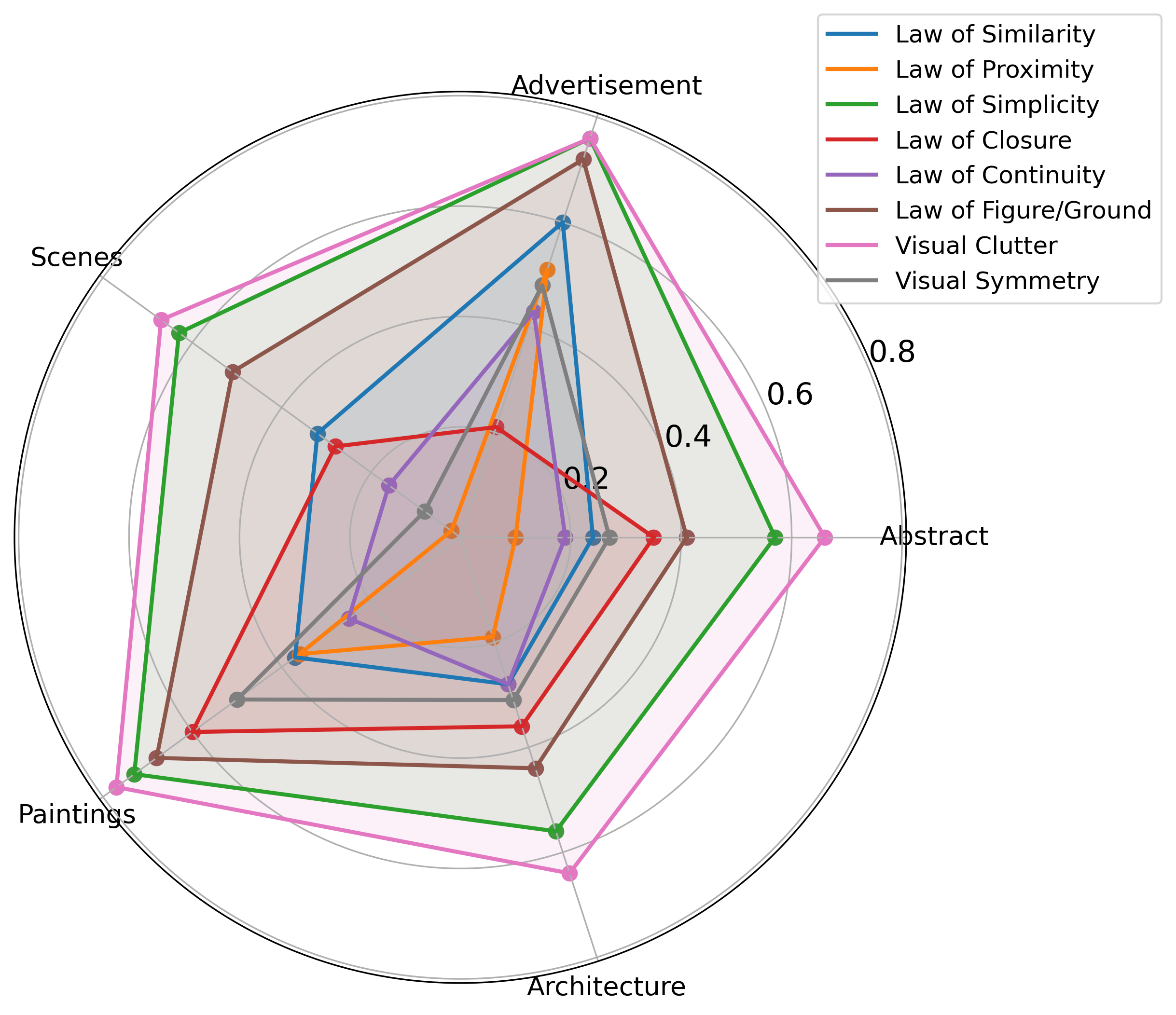} 
        \caption{SROCC scores for IC9600 dataset}
        \label{fig:srocc_ic9600}
      \end{subfigure}%
      \hspace*{\fill}%
  \caption{Results summary for SAVIOAS and IC9600 dataset. Detailed results can be found in tables \ref{table:savoias_results} and \ref{table:ic9600_results} in the appendix.}
  \label{fig:radar_plots}
\end{figure}

\section{Results \& Discussion}


Figure \ref{fig:radar_plots} shows the correlation between scores obtained through pairwise comparisons of images using explainable measures of visual perception and human annotations for complexity. We find that scores for visual clutter and law of simplicity based comparisons have highest correlation with human annotations. This demonstrates that most of the human annotators used in datasets like SAVOIAS and IC9600 for comparing and analyzing visual perception in images are biased towards the notions of visual clutter and visual simplicity causing a negligence of other reasonings proposed in psychology and cognitive science literature. We further find that these measures are consistent across different categories. These results further corroborate with the findings of prevailing studies that emphasize the role of clutter \cite{yu2014modelingvisualclutter} and simplicity \cite{feldman2016simplicityprinciple} principles in human judgment for aesthetic appeal or visual complexity of visual stimuli. We further find that several principles impact various categories differently. For example, law of closure does not correlate well with human labels from advertisement category but have a much higher correlation for categories like suprematism and paintings. We observe consistent performance in the Advertisement category across datasets, highlighting the method's robustness to biases and varying annotation strategies. Certain optimizations like Bradley-Terry and matrix completion can be applied to reduce the computation overload by reducing number of pairwise comparisons. Figure \ref{fig:MLLM_sample} illustrates a qualitative comparison example for interior design images. 

\section{Applications \& Future Work}
Some potential applications include optimizing user interfaces and experiences using specific principles from cognitive science. Our framework can aid in assistive technologies and guide the development of accessible digital environments for individuals with neurodiverse conditions like dyslexia or ADHD since visual complexity directly impacts cognitive accessibility. Content creators and designers can leverage our approach to generate explainable feedback on visual balance, clutters and symmetry in digital artwork or while evaluating several templates/figmas to improve visual appeal. Moreover, MLLMs can refine search result presentation by assessing the influence of visual complexity on user decisions thereby enhancing product ranking algorithms to balance aesthetics with information clarity. Human-AI collaborations can help refine MLLM reasoning further, thereby improving the overall quality of XAI (Explainable AI) systems before deploying them in real-world tasks.   
This study also hints at the tendency of human-annotated datasets to overemphasize such principles of clutter and simplicity in judgement of visual complexity thereby influencing overall correlations. There is room for improvement in understanding the sensitivity of MLLM methods to the quality and design of the prompts. Furthermore, evaluating the performance of different MLLMs like GPT 4v \cite{openai2024gpt4technicalreport}, Gemini Pro \cite{geminiteam2024gemini15unlockingmultimodal}, etc. could provide new complementary insights on the explainable perception of visual content while also providing a useful benchmark test to evaluate the multimodal contextualizing capabilities of MLLMs.

\bibliographystyle{ACM-Reference-Format}
\bibliography{sample-base}

\section{Appendix}

\begin{table}[h!]
\centering
\resizebox{\textwidth}{!}{
\begin{tabular}{|l|c|c|c|c|c|c|c|c|c|c|c|c|}
\hline
\multirow{2}{*}{\textbf{Law/Concept}} & \multicolumn{2}{c|}{\textbf{Advertisement}} & \multicolumn{2}{c|}{\textbf{Interior Design}} & \multicolumn{2}{c|}{\textbf{Objects}} & \multicolumn{2}{c|}{\textbf{Scenes}} & \multicolumn{2}{c|}{\textbf{Suprematism}} & \multicolumn{2}{c|}{\textbf{Visualizations}} \\
\cline{2-13}
 & \textbf{PLCC} & \textbf{SROCC} & \textbf{PLCC} & \textbf{SROCC} & \textbf{PLCC} & \textbf{SROCC} & \textbf{PLCC} & \textbf{SROCC} & \textbf{PLCC} & \textbf{SROCC} & \textbf{PLCC} & \textbf{SROCC} \\
\hline
\textbf{Law of Similarity} & $0.49$ & $0.52$ & $0.69$ & $0.72$ & $0.52$ & $0.51$ & $0.49$ & $0.36$ & $0.65$ & $0.63$ & $0.5$ & $0.5$ \\
\textbf{Law of Proximity} & $0.41$ & $0.39$ & $0.66$ & $0.70$ & $0.40$ & $0.39$ & $0.20$ & $0.07$ & $0.20$ & $0.19$ & $0.5$ & $0.51$ \\
\textbf{Law of Simplicity} & $0.65$ & $\boldsymbol{0.71}$ & $\boldsymbol{0.73}$ & $\boldsymbol{0.76}$ & $0.73$ & $0.77$ & $0.65$ & $0.63$ & $0.79$ & $0.80$ & $0.62$ & $0.63$ \\
\textbf{Law of Closure} & $0.02$ & $0.03$ & $0.24$ & $0.34$ & $0.27$ & $0.28$ & $0.34$ & $0.31$ & $0.55$ & $0.52$ & $0.36$ & $0.41$ \\
\textbf{Law of Continuity} & $0.35$ & $0.30$ & $0.70$ & $0.72$ & $0.22$ & $0.22$ & $0.37$ & $0.29$ & $-0.05$ & $-0.08$ & $0.39$ & $0.42$ \\
\textbf{Law of Figure/Ground} & $0.56$ & $0.67$ & $0.68$ & $0.73$ & $0.55$ & $0.65$ & $0.49$ & $0.45$ & $0.56$ & $0.58$ & $0.56$ & $0.59$ \\
\textbf{Visual Clutter} & $\boldsymbol{0.67}$ & $0.69$ & $0.65$ & $0.64$ & $\boldsymbol{0.77}$ & $\boldsymbol{0.79}$ & $\boldsymbol{0.70}$ & $\boldsymbol{0.65}$ & $\boldsymbol{0.82}$ & $\boldsymbol{0.82}$ & $\boldsymbol{0.67}$ & $\boldsymbol{0.66}$ \\
\textbf{Visual Symmetry} & $0.45$ & $0.47$ & $0.61$ & $0.63$ & $0.04$ & $0.08$ & $0.00$ & $-0.04$ & $0.61$ & $0.60$ & $0.42$ & $0.41$ \\
\hline
\end{tabular}}
\caption{SAVOIAS dataset: Correlation Coefficients of predicted visual complexity measures and groundtruth obtained through crowdsourcing.}
\label{table:savoias_results}
\end{table}

\begin{table}[h!]
\centering
\resizebox{\textwidth}{!}{
\begin{tabular}{|l|c|c|c|c|c|c|c|c|c|c|}
\hline
\multirow{2}{*}{\textbf{Law/Concept}} & \multicolumn{2}{c|}{\textbf{Abstract}} & \multicolumn{2}{c|}{\textbf{Advertisement}} & \multicolumn{2}{c|}{\textbf{Scenes}} & \multicolumn{2}{c|}{\textbf{Paintings}} & \multicolumn{2}{c|}{\textbf{Architecture}} \\
\cline{2-11}
 & \textbf{PLCC} & \textbf{SROCC} & \textbf{PLCC} & \textbf{SROCC} & \textbf{PLCC} & \textbf{SROCC} & \textbf{PLCC} & \textbf{SROCC} & \textbf{PLCC} & \textbf{SROCC} \\
\hline
\textbf{Law of Similarity} & $0.35$ & $0.24$ & $0.55$ & $0.60$ & $0.36$ & $0.32$ & $0.49$ & $0.37$ & $0.35$ & $0.28$ \\
\textbf{Law of Proximity} & $0.17$ & $0.10$ & $0.51$ & $0.51$ & $0.10$ & $0.02$ & $0.45$ & $0.36$ & $0.27$ & $0.19$ \\
\textbf{Law of Simplicity} & $0.52$ & $0.57$ & $0.70$ & $\boldsymbol{0.76}$ & $0.61$ & $0.63$ & $0.74$ & $0.73$ & $0.52$ & $0.56$ \\
\textbf{Law of Closure} & $0.32$ & $0.35$ & $0.18$ & $0.21$ & $0.18$ & $0.28$ & $0.45$ & $0.60$ & $0.33$ & $0.36$ \\
\textbf{Law of Continuity} & $0.19$ & $0.19$ & $0.44$ & $0.43$ & $0.17$ & $0.16$ & $0.35$ & $0.25$ & $0.30$ & $0.28$ \\
\textbf{Law of Figure/Ground} & $0.37$ & $0.41$ & $0.64$ & $0.72$ & $0.46$ & $0.51$ & $0.61$ & $0.68$ & $0.39$ & $0.44$ \\
\textbf{Visual Clutter} & $\boldsymbol{0.64}$ & $\boldsymbol{0.66}$ & $\boldsymbol{0.76}$ & $\boldsymbol{0.76}$ & $\boldsymbol{0.68}$ & $\boldsymbol{0.67}$ & $\boldsymbol{0.77}$ & $\boldsymbol{0.77}$ & $\boldsymbol{0.62}$ & $\boldsymbol{0.64}$ \\
\textbf{Visual Symmetry} & $0.26$ & $0.27$ & $0.46$ & $0.48$ & $0.08$ & $0.08$ & $0.50$ & $0.50$ & $0.31$ & $0.31$ \\
\hline
\end{tabular}}
\caption{IC9600 dataset: Correlation Coefficients of predicted visual complexity measures and groundtruth obtained through crowdsourcing.}
\label{table:ic9600_results}
\end{table}










\end{document}